\documentclass[12pt]{iopart}
\usepackage{amsfonts,latexsym,cite}

\newcommand{\be}{\begin{equation}} 
\newcommand{\ee}{\end{equation}}
\newcommand{\ba}{\begin{eqnarray}} 
\newcommand{\ea}{\end{eqnarray}}

\def\nn{\nonumber} \def\dag{\dagger}

\def\lf{\left}

\def\ri{\right}
\newcommand\ZZ{{\mathbb Z}}

\newcommand\eq{\begin{equation}}
\newcommand\en{\end{equation}}
\newcommand{\fract}[2]{{\textstyle\frac{#1}{#2}}}
\newcommand\hf{\fract{1}{2}}

\vspace{6cm}

\def\a{\alpha}
\def\b{\beta}
\def\e{\epsilon}

\def\beq{\begin{equation}}
\def\eeq{\end{equation}}
\def\bea{\begin{eqnarray}}
\def\eea{\end{eqnarray}}
\def\ba{\begin{array}}
\def\ea{\end{array}}

\def\lt{\left}
\def\rt{\right}

\def\nn{\nonumber} \def\dag{\dagger}
\def\e{\epsilon}

\def\a{\alpha}
\def\b{\beta}

\begin{document}

\paper
{Some spectral equivalences between Schr\"odinger operators
} 

\author{C Dunning\footnote[1]{t.c.dunning@kent.ac.uk}, 
K E Hibberd\footnote[2]{keh@maths.uq.edu.au} and 
J Links\footnote[3]{jrl@maths.uq.edu.au}}  
\address{$\dag$ Institute of Mathematics, Statistics and Actuarial
Science, The University of Kent, United Kingdom} 
\address{$\ddag\S$ Centre for Mathematical Physics, 
School of Physical Sciences, The University of
Queensland, Brisbane, 4072, Australia}

\begin{abstract}
Spectral equivalences of the quasi-exactly solvable  sectors of  two classes  of
Schr\"odinger operators are 
established, using Gaudin-type Bethe ansatz equations. In some instances the results can be extended leading to full isospectrality.   In this manner we obtain 
equivalences between $PT$-symmetric problems and Hermitian
problems. 
We also find equivalences  between  some classes of Hermitian operators.  
\end{abstract}
\pacs{02.30.Hq, 0.230.Ik, 03.65.Ge} 

\maketitle  


\section{Introduction}

In recent years there has been substantial activity studying the
relationships between integrable systems, which can be exactly solved
via Bethe ansatz methods, and the spectra of differential equations
\cite{DTa,BLZa,Sa,DTb}. See the review article
\cite{ddt_review} for further references.  This has lead to a deep
understanding 
of the spectral properties of certain Schr\"odinger operators which are
not Hermitian, but possess the more general property of
$PT$-symmetry~\cite{BB,BBM}. 
An initially surprising result was the establishment of
reality for the energy levels of particular non-Hermitian operators
with $PT$-symmetry \cite{DDT3,DDTs}. An
alternative approach to prove 
 reality  of a $PT$-symmetric operator  
is to construct an equivalent Hermitian Hamiltonian~\cite{bbj,sgh,mos,mos1}. 
Such a construction has proven difficult
 and only a few explicit (non-perturbative) results are known 
 \cite{and,bg,swanson,jm,bm,jones_lee,Bender:2008uu}. 
For a detailed review of the field we refer to \cite{B1,B2,ddt_review}.

Our goal here is to further expose spectral equivalences between
Schr\"odinger operators which are quasi exactly solvable (QES)
\cite{Tur,Ush}, i.e. operators for which part of the spectrum can be
determined algebraically. We will consider two cases, viz. one
associated with sextic potentials with an angular momentum like term,
and another where the potentials are expressed in terms of hyperbolic
functions. For both cases the starting point is to begin with a
Hamiltonian which admits two exact Bethe ansatz solutions. Here, the
Bethe ansatz solutions are of the Gaudin (additive) form with finitely
many roots, so our approach is different from the cases
\cite{DTa,BLZa,Sa,DTb,DDT3,DDTs} 
for (multiplicative) Bethe ansatz
equations with an infinite number of roots.
 Each solution can be mapped to the QES sector of a one-dimensional Schr\"odinger operator, and these turn out to have different potentials. Since the QES eigenvalues of the Schr\"odinger operators are the same as the Hamiltonian, this establishes a spectral equivalence at the level of the QES sectors.       

For the sextic case our results show there is a spectral equivalence
of the QES levels between certain Hermitian and a $PT$-symmetric
potentials. This provides a starting point to determine equivalences
in a more general context, which we also discuss. In some instances we
can establish these equivalences rigorously, using the techniques of
Bender-Dunne polynomials \cite{BD} and Darboux-Crum \cite{D,C}
transformations. In other cases we give conjectures which we find are
supported by numerical calculation of the spectra.   

In the last section of the paper we study a second class of Hamiltonians
from which we can determine spectral equivalences of the QES levels
for hyperbolic potentials. Here we find that the equivalence exists
between two Hermitian potentials or alternatively, by use of a unitary
transformation, two $PT$-symmetric potentials. This contrasts the case
of the sextic potentials where the equivalence is between Hermitian
and $PT$-symmetric potentials.  For the normalisable cases, we
obtain a complete spectral equivalence by showing
that the two potentials are supersymmetric partners \cite{Wit,Shif}.

\section{Spectral equivalences in sextic potentials}


We begin by considering a class of Hamiltonians given by
\begin{eqnarray}
H_p=A_p S_p^z + B_p(d_p^\dagger S_p^-+\xi_p d_p S_p^+),
\label{ham0}
\end{eqnarray} 
where $\{d_p,d_p^\dagger\}$ are boson operators and $\{S_p^z,\,S_p^+,S_p^-\}$ are $su(2)$ operators, 
while $\xi_p=\pm 1$ is a discrete variable. Through use of the algebraic Bethe ansatz (ABA) method in the {\it quasi-classical limit} (see for example \cite{lzmg}), 
the energies and corresponding Bethe ansatz equations (BAE) are found to be
\bea 
&&E_p=A_p( M_p+\kappa_p)+\xi_p B_p\sum_{j=1}^{M_p} v_p^{(j)}  \label{energy} ,\\
&&\frac{2\kappa_p}{v_p^{(j)} }+\xi_p v_p^{(j)} +\frac{A_p}{B_p}=\sum_{k\neq j}^{M_p}\frac{2}{v_p^{(k)} -v_p^{(j)} }\label{BAE}.
\eea
The parameter $\kappa_p$ is determined by the reference state $\left|\phi_p\right>$ which satisfies 
\begin{eqnarray*}
S_p^z\left|\phi_p\right>=\kappa_p\left|\phi_p\right>,~~~~S_p^-\left|\phi_p\right>=0~~~~d_p\left|\phi_p\right>=0,
\label{vac}
\end{eqnarray*}
and the eigenstates have the form
\bea
\left|\{  v_p^{(k)} \}\right>=\prod_{j=1}^{M_p}(v_p^{(j)} d_p^\dagger+S^+_p)\left|\phi_p\right> .
\label{states}
\eea 


\subsection{Equivalences of QES sectors}

Next we consider the Hamiltonian 
\bea 
H=\epsilon(n_a-n_b-n_c)+\Omega(a^\dagger b c + a b^\dagger c^\dagger)\label{ham}
\eea
where $\{\alpha,\alpha^\dagger:\alpha=a,b,c\}$ are canonical boson operators with the usual number operators $n_\alpha=\alpha^\dagger \alpha$. 
It is straightforward to verify that this Hamiltonian commutes with the conserved operators
$ N=2n_a+n_b+n_c,~K=n_b-n_c. $ 
To make a connection with (\ref{ham0}), we make the following two realisations of  $su(2)$ operators, 
\bea 
S^+_1= a^\dagger c,~~~~S^-_1=a c^\dagger,~~~~S_1^z=\frac{1}{2}(n_a-n_c),~~~~ \nonumber \\  
S^+_2= -b^\dagger c^\dagger,~~~~S^-_2=b c,~~~~S_2^z=\frac{1}{2}(n_b+n_c+1)~~~~ 
\eea
and set 
$$ d^\dagger_1=b^\dagger,\qquad d^\dagger_2=a^\dagger . $$
We may now express the Hamiltonian (\ref{ham}) as
\bea 
H&=&H_1 -\frac{\epsilon}{4}(N+3K) = H_2 +\frac{\epsilon}{2}(N+3I)\label{ham12}
\eea
where 
\begin{eqnarray}
A_1=-A_2=3\epsilon,\qquad B_1=B_2=\Omega,\qquad \xi_1=-\xi_2=1.
\label{condition0}
\end{eqnarray}
Hereafter we set $\Omega=1$.

For the reference states we choose $\left|\phi_p\right>
=\left|\phi(q_p)\right>$  
\begin{eqnarray*} 
&& \left|\phi(q_p)\right> = \frac{1}{\sqrt{q_p!}}(c^\dagger)^{q_p} \left|0\right>, 
\end{eqnarray*}
which leads to the values 
\begin{eqnarray}
\kappa_1= -q_1/2 ,\,\qquad \kappa_2= (q_2+1)/2.
\label{condition1}
\end{eqnarray} 
Given a solution for $H$ with the $M_1$ Bethe roots $\{v_1^{(j)}\}$ associated with $H_1$ and $q_1$, we need to determine the 
relationship to the solution of the $M_2$ Bethe roots 
$\{ v_2^{(j)}\}$ for $H_2$ with $q_2$. 
{}From the form of the eigenstates (\ref{states}) we can deduce the values of the conserved operators
\begin{eqnarray*} 
N&=& M_1+q_1 =  2M_2+q_2, \\
K&=&  M_1-q_1 =  -q_2. 
\end{eqnarray*} 
Solving this gives
\begin{eqnarray}
q_1=M_2+q_2,~~~~~M_1=M_2. 
\label{condition2}
\end{eqnarray}
This imposes the restriction $M_1\leq q_1$. (Note that for $M_1>q_1$ the eigenstate (\ref{states}) vanishes.) 
For convenience set $q_2 =q$ and $M_2=M$.  If $E$ is the energy
corresponding to the Hamiltonian (\ref{ham}) then through (\ref{ham12}) we have
\bea 
E&=&E_1 -\frac{\epsilon}{2}( M-q)  \nn\\
&=& E_2 +\frac{\epsilon}{2}(2M+q+3).\label{en}
\eea

%

At this point we remark that the energy expression (\ref{energy}) and associated Bethe ansatz equations (\ref{BAE}) have precisely the form of the exact solution for the QES Schr\"odinger operator with sextic potential \cite{Ush}. Explicitly, 
\beq
-\psi_p''(x) +V_p(x) \psi_p(x) =0,
\label{sch}
\eeq
$$V_p (x) = 4E_p + x^6+ 2A_p  \xi_p x^4 +
  [ \xi_p( 4M+4\kappa_p +2)+ A_p^2 ] x^2 +
\frac {(2\kappa_p -1/2)( 2\kappa_p-3/2) }{x^2} ,$$
for the functions
\bea
\psi_p (x)  &= & x^{2\kappa_p -\frac 12} \exp{\lt[ \frac{x^2}2 \lt(A_p +  \frac{\xi_px^2}2\rt)\rt] } Q_p(x),  \nn\\
Q_p(x) &=& \prod^{M_p}_{j=1} \lt(x^2 -  v_p^{(j)}  \rt).\nn
\eea
Setting $\hat E = -4( E+\e/2(M-q) )$ and using the relations (\ref{condition0},\ref{condition1},\ref{condition2},\ref{en}) we can write 
\begin{eqnarray*}
-\psi_p''(x) +V_p(x) \psi_p(x) =\hat E ~\psi_p(x) 
\end{eqnarray*}
\begin{eqnarray}
V_{1}(x) &=&  x^6+ 6\e x^4  + x^2  [ 2( M-q +1)+ 9\e^2]     \nn\\
&&+\frac {(M+q +1/2)( M+q+3/2) }{x^2}   \label{v1} \\
\nn\\
V_{2}(x) &=&   x^6+ 6\e x^4  + x^2  [ - 4M-2(q+2)+ 9\e^2]- 6\e(M+1)  \nn\\
&& +\frac {(q+1/2)( q-1/2) }{x^2}  \label{v2}
\end{eqnarray}
and see that the $p=1$ case shares the same QES spectrum as the $p=2$ case.

Before proceeding further some remarks are required. Consider the general potential  
 \eq
V = x^6 +2\delta x^4 + (\delta^2 +\alpha)x^2
+\frac{l(l+1)}{x^2} +C 
\label{gen}
\en
where $\delta, \alpha $ and $l$ are real parameters and $C$ is
 a real constant. Usually   we  require the 
 wavefunction $\psi(x)$  to be square integrable along the
 positive real axis while behaving at the origin as  $\psi|_{x\to
  0} \sim x^{l+1}$. 
If $l >  -1/2$ this defines a Hermitian problem and 
so we will use the subscript $H$ to indicate this type of radial 
Schr\"odinger problem.  We can alternatively consider $PT$-symmetric boundary
conditions \cite{BB,BBM}, denoted by the subscript $PT$.  In this case, we
require the 
wavefunction to be square 
integrable along a contour in the complex plane ${\cal C}$ which
 starts and ends at $|x| =  \infty$ within the Stokes sectors 
defined as wedges in the complex plane of open angle $\pi/4$, centered 
at angles $-\pi/4$ and $\pi/4$.   This contour should also avoid the
origin whenever $l(l+1) \ne 0$. 
In view of this we see that the $p=1$ case of (\ref{sch}) corresponds to
the $PT$-symmetric case while $p=2$ is the Hermitian case.

We mention also that the change of variable $\epsilon \rightarrow
-\epsilon$ is equivalent,  
up to a unitary transformation $a^\dagger\rightarrow -a^\dagger$, to the mapping $H \rightarrow -H$. 
Hence we also have the equivalence that the QES spectrum of $V_p(x;\epsilon)$ is the negative of the 
QES spectrum of $V_p(x;-\epsilon)$ for both $p=1,2$; this is the
anti-isospectral duality of \cite{kuw}.



\subsection{ Equivalence beyond the QES spectrum}


In this subsection we will describe how the spectral equivalence
obtained in the last section between the QES eigenvalues of a
$PT$-symmetric and 
a Hermitian Schr\"odinger problem is in fact a complete spectral
equivalence.   

If we set  $\alpha=\alpha_J= -(4J+1+2l)$ and $C=-2\delta J$ in
(\ref{gen}) then the 
results of the 
last section prove that the Schr\"odinger operators with potentials
\bea
V_{\rm H} &=& x^6 +2\delta x^4 + (\delta ^2  -(4J+1+2l))x^2
+\frac{l(l+1)}{x^2} -2\delta J,\nn\\
&& \mbox{\hspace{2cm}}
\quad \psi|_{x\to
  0} \sim x^{l+1} \nn \\
V_{\rm PT} &=& x^6 +2\delta x^4 +(\delta^2+2J-1-2l
 )x^2
+\frac{(l+J)(l+J+1)}{x^2}
\label{hpt}
\eea
 are both quasi-exactly solvable and the $J=M+1$ exactly-known 
eigenvalues coincide.

 The QES spectral equivalence (\ref{hpt}) at $\delta=0$ was discovered
and proven  \cite{DDT3} 
via the ODE/IM correspondence \cite{DTa,BLZa}. 
In fact, the 5th spectral equivalence of
\cite{DDT3} makes a  stronger statement: it says that the full
spectrum of the $PT$-symmetric and Hermitian problems (\ref{hpt})  are
  isospectral,  not just the 
QES levels.   Moreover, the equivalence also holds away from the special
QES points $\alpha_J.$    Numerically, we find this equivalence also extends
to the case when $\delta\neq 0$. 
We conjecture that the most general equivalence
is that 
\bea
V_{\rm H} &=& x^6 +2\delta x^4 + (\delta ^2 +\alpha)x^2
+\frac{l(l+1)}{x^2} ,\quad \psi|_{x\to
  0} \sim x^{l+1}
\label{hpth}
\eea
is isospectral to
\eq
\fl V_{\rm PT} = x^6 +2\delta x^4 +(\delta^2-\hf
 (\alpha+6l+3))x^2
+\frac{(2l+3-\alpha)(2l-1-\alpha)}{16x^2}
+\frac{\delta}{2} (\alpha + 1 + 2l).
\label{hpt2}
\en
The approach used in the
previous section does not allow us to prove this statement away  from 
the QES sector for the special points $\alpha_J$.  
It would be interesting to generalise the approach of  \cite{DDT3} in
order to obtain a rigorous proof of this statement.   Numerical
confirmation of the spectral equivalence (\ref{hpth}) and (\ref{hpt2})
for values of $(\delta,\alpha,l)$ away from the QES points is shown in  
table~\ref{tb1}. 

\begin{table}[tb]
\begin{center}
\begin{tabular}{|c|c|c|}
\hline $n$ & $E_n$ Hermitian & $E_n$ PT \\   
\hline
0 &  7.17030615 &  7.17030616 \\
1 &  19.5220637 &  19.5220637\\
2 &  35.2744653  & 35.2744654 \\ 
3 &  53.7929337 &  53.7929339 \\
4 &  74.7062464  & 74.7062466 \\  
\hline 
\end{tabular}
\end{center}
\caption{\label{tb1} \footnotesize The spectrum of (\ref{hpth}) and
  (\ref{hpt2})  
  with  $\delta=0.2,\alpha=0.31, l=0.54$.   }  
\end{table}


\subsection{Further spectral equivalences}  


Here we briefly comment that the further spectral equivalences
obtained in \cite{DDT3} for  (\ref{gen}) with $\delta =0$ can also be
generalised to the problem with $\delta \ne 0$. 

The second equivalence of \cite{DDT3} relating the spectrum of two
Hermitian sextic problems  
is easily generalised using the results of the last section. 
The $PT$-symmetric problem (\ref{hpt2}) is invariant under
$(\delta,\alpha,l)\to (\delta, (6l+3-\alpha)/2 ,
(\alpha+2l-1)/4 )$
whereas the same transformation on the Hermitian case (\ref{hpth})
leads to a different Hermitian problem.   Rewriting, 
 we obtain a full spectral equivalence between two radial problems: 
\bea
V_{\rm H}^1 &=& x^6 +2\delta x^4 + (\delta^2+\alpha)x^2 +\frac{ l(
  l+1)}{x^2} - 
\frac{\delta}{4}(1+2l-\alpha) ,
\nn\\
&&  \mbox{\hspace{9cm}}
  \psi|_{x\to
  0} \sim x^{l+1} \nn \\
V_{\rm H}^2 &=& x^6 +2\delta x^4 + (\delta^2+\frac 1 2 (3-\alpha+6 l))x^2
+\frac{(2l-1+\alpha)(2l+3+\alpha)}{16x^2} \nn \\
&&+  
\frac{\delta}{4}(1+2l-\alpha) ,\qquad \qquad  \quad \quad \psi|_{x\to
  0} \sim x^{(\alpha +2l-1)/4+1}.
\label{spec2}
\eea
This equivalence can be proven in terms of Bender-Dunne
polynomials \cite{BD}.   Set 
 \eq
 \psi(x,E,\delta,\alpha,l)= e^{-x^4/4 -\delta x^2 /2} \,x^{l+1} \sum_{n
=0}^\infty 
\lf (\frac {-1}{{\ }4} \ri )^n \frac{P_n(E,\delta,\alpha,l)}{n! \Gamma
 (n+l+3/2)} x^{2n}. 
\label{psi}
 \en
To satisfy a radial  ODE with general potential (\ref{gen}) the
polynomials $P_n$ 
must satisfy the three-term recursion relation
\bea
P_n(E) &=& (E+C-\delta (2l+4n-1))P_{n-1}(E) \nn\\
&&+
16(n-1)(n+(\alpha+2l-3)/4)(n+l-1/2)P_{n-2} \nn 
\eea
with $P_0(E)=1, P_0(E) = E$.  
The wavefunction  $\psi$ is an   
everywhere-convergent series for
 all $l \neq -n-3/2 , n \in \ZZ$ for arbitrary $\alpha$, with the
 required behaviour at the origin built-in: that is,
 $\psi(x)\sim x^{l+1}$ at $x=0$. 
For generic values of $\alpha$, the wavefunction $\psi$ will  be an
infinite series. 
However, whenever $\alpha=-(4J+2l+1)$ for positive integer $J$,
the series  truncates and the zeros of $P_J(E)$ are the QES
eigenvalues \cite{BD}. 
 The relevant point here is that 
the recursion relation is  invariant under 
\eq
\alpha \to  (6l+3-\alpha)/2 \quad,\quad   l \to
(\alpha+2l-1)/4 \quad,\quad C\to C-\delta(1+2l-\alpha)/2, \quad 
\en
giving rise, with $C=-\delta(1+2l-\alpha)/4$, to the spectral
equivalence (\ref{spec2}).

One more spectral equivalence may be obtained, either via the 
Bender-Dunne polynomials as in \cite{DDT3} or  using Darboux-Crum
transformations \cite{D,C}, generalising   the third spectral equivalence
of \cite{DDT3}.  Setting $\alpha_J=-(4J+2l+1)$ we take the
Hermitian QES problem  and using a Darboux-Crum transformation
remove all of the QES 
 levels leaving the rest of the spectrum in place.  
The resulting potential is once again exactly the same as
the original form modulo a change in the parameters.  That is, the potential 
\bea
 V_{\rm H}(x) &=& x^6 + 2 \delta x^4 + (\delta^2 -(4J+2l+1)) x^2 +
 \frac{l(l+1)}{x^2 }  ,\quad\nn\\
&& \mbox{\hspace{2cm}}
 \psi|_{x\to 0} \sim x^{l+1} \label{hermq}
 \eea
is isospectral to 
\bea
 V_{\rm H}(x)&=& x^6 + 2 \delta x^4 + (\delta^2 +(2 J-2 l-1)) x^2 +
 \frac{( l + J)( l +  J+1)}{x^2 }\nn\\
&&  +2\delta J  ,\quad \mbox{\hspace{2cm}}
 \psi|_{x\to 0} \sim x^{ l+J}
\eea
except for the first $J$ QES levels  of (\ref{hermq}).    The result is
unexpected because the intermediate potentials, found by
removing one energy level at a time, are in general singular potentials.
  Note that this result can also be obtained  from the cubic case  
 of the type A ${\cal N}$-fold supersymmetry of \cite{nfold}.

Finally, combining the above equivalences, we find  that 
the QES problem 
\eq
V_{\rm H} = x^6 +2\delta x^4 + (\delta ^2 - (4J+2l+1))x^2 +
\frac{l(l+1)}{x^2} ,\quad
 \psi|_{x\to 0} \sim x^{ l+1}
\label{vh}
\en
is isospectral to
\eq
V_{\rm PT} = x^6 +2\delta x^4 + (\delta ^2 - (4J+2l+1))x^2 +
\frac{l(l+1)}{x^2} 
\label{vpt}
\en
except for the QES eigenvalues. Numerical confirmation is shown in table~\ref{tb2}. 
\begin{table}[tb]
\begin{center}
\begin{tabular}{|c|c|c|}
\hline $n$ & $E_n$ Hermitian & $E_n$ PT \\   
\hline
0  &-11.0798088 &  30.1033297 \\ 
1  &  1.45911359 &  48.4085577\\
2 &  14.4686962 &  69.0856538\\
3 &  30.1033293 &  91.8988711  \\ 
4 &  48.4085576 &  116.668373  \\
\hline 
\end{tabular}
\end{center}
\caption{\label{tb2} \footnotesize The spectrum of 
   (\ref{vh}) and 
  (\ref{vpt})  with   $\delta=0.2, J=3, l=0.54$.   }   
\end{table}


\section{Spectral equivalences in hyperbolic potentials}  

Now we move on to examine another case in which a spectral equivalence can be established for potentials expressed in terms of hyperbolic functions.
Here we consider the following Hamiltonian
\begin{eqnarray}
H=\epsilon(n_1+n_2-n_3-n_4)+g(n_1n_3+n_2n_4 + a_1^\dagger a_2^\dagger a_3 a_4 + a_1 a_2 a_3^\dagger a_4^\dagger),
\label{ham3}
\end{eqnarray}
where the $\{a_j,\,a_j^\dagger\, |j=1,2,3,4\}$ are canonical boson operators
and $n_j=a_j^\dagger a_j$. We note that the unitary transformation 
\begin{eqnarray}
a_1^\dagger \longleftrightarrow a_3^\dagger ,~~~
a_2^\dagger \longleftrightarrow a_4^\dagger 
\label{ut}
\end{eqnarray}
is equivalent to the change of variable 
\begin{eqnarray}
\epsilon \rightarrow -\epsilon.
\label{cov1}
\end{eqnarray}
 
We make the following assignment of the $su(2)$ operators with a central extension:
\begin{eqnarray*}
&&S_1^+=a_1^\dagger a_2^\dagger,~~~~~S_1^-=-a_1 a_2,~~~S_1^z=\frac{1}{2}(n_1+n_2+1),
~~~K_1=\frac{1}{2}(n_1-n_2), \\
&&S_2^+=-a_3^\dagger a_4^\dagger,~~\,S_2^-=a_3 a_4,~~~~~S_2^z=\frac{1}{2}(n_3+n_4+1),~~~
K_2=\frac{1}{2}(n_3-n_4), \\ 
&&S_3^+=a_1^\dagger a_4,~~~~~S_3^-=a_4^\dagger a_1,~~~~~S_3^z=\frac{1}{2}(n_1-n_4),~~~~~~~~
K_3=\frac{1}{2}(n_1+n_4), \\
&&S_4^+=a_3^\dagger a_2,~~~~~S_4^-=a_2^\dagger a_3,~~~~~S_4^z=\frac{1}{2}(n_3-n_2),~~~~~~~~
K_4=\frac{1}{2}(n_2+n_3), \\
\end{eqnarray*}
which for $j=1,2,3,4$ satisfy the  commutation relations: 
\begin{eqnarray*}
[S_j^z,\,S_j^\pm]=\pm S_j^\pm,~~~~[S_j^+,\,S_j^-]=2S_j^z,~~~~[K_j,\,S^z_j]=[K_j,\,S_j^\pm]=0. 
\end{eqnarray*}
We will also need the corresponding Casimir invariants
$$C_j=S^+_j S^-_j+ S_j^z(S_j^z-I) . $$
In terms of these operators we can express the Hamiltonian (\ref{ham3}) as 
\begin{eqnarray}
H&=& H_{\a}-g(C_1+C_2-(S^z_1+S^z_2)^2-2K_1K_2+2(S_1^z+S_2^z)-\frac{1}{2}I ) \label{ham1}\\
&=&H_{\b}-g(C_3+C_4-(S^z_3+S^z_4)^2-2K_3K_4+S^z_3+S^z_4 ). \label{ham2}
\end{eqnarray}
where 
\begin{eqnarray*}
H_{\a}&=& 2\epsilon(S_1^z-S_2^z)+g(S^+_1S^-_1+S^+_2S^-_2+S^+_1S^-_2+S^+_2S^-_1),\nn\\
H_{\b}&= &2\epsilon(S_3^z-S_4^z)+g(S^+_3S^-_3+S^+_4S^-_4+S^+_3S^-_4+S^+_4S^-_3).
\end{eqnarray*}
We recognise from \cite{lzmg} (see eq. (69)) that $H_{\a}, H_{\b}$ are Bethe ansatz solvable. 
Since the additional terms appearing in (\ref{ham1}) commute with $H_{\a}$, 
we can extend it to a solution for $H$. Likewise, since the terms in (\ref{ham2}) commute with $H_{\b}$, 
we can also obtain a second solution from this expression.  

To obtain the Bethe ansatz solutions we need to identify a reference state $\left|\phi\right>$ which satisfies
\begin{eqnarray*}
S_j^-\left|\phi\right>=0,~~~~S^z_j\left|\phi\right>=-s_j\left|\phi\right>
\end{eqnarray*}
for some scalars $s_j$; i.e. $\left|\phi\right>$ is a lowest weight state for all realisations of the $su(2)$ algebra. 
The choices 
\begin{eqnarray}
\left|\phi(p_\sigma,q_\sigma)\right>=\frac{1}{\sqrt{p_\sigma!q_\sigma!}}(a_2^\dagger)^{p_\sigma}(a_4^\dagger)^{q_{\sigma}}\left|0\right>,
\label{vac2}
\end{eqnarray}
$\sigma=\alpha,\,\beta$
satisfy this condition with 
\begin{eqnarray*}
s_1=-\frac{p_\a+1}{2},~~~s_2=-\frac{q_\a+1}{2},~~~s_3=\frac{q_\b}{2},~~~s_4=\frac{p_\b}{2}.
\end{eqnarray*}

Now we can write down the exact solution from \cite{lzmg}. 
For the representation (\ref{ham1}) we have that the energies 
of $H_{\a}$ are given by 
\begin{eqnarray}
E_{\a}=(p_\a-q_\a)\epsilon  -2\sum_{j=1}^{M_\a} v_\a^{(j)} 
\end{eqnarray}
where the $\{v_\a^{(j)}\}$ are solutions of the BAE
\begin{eqnarray}
\frac{2}{g}-\frac{p_\a+1}{ v_\a^{(j)} +\epsilon}-\frac{q_\a+1}{v_\a^{(j)} -\epsilon}=\sum_{k\neq j}^M\frac{2}{v_\a^{(j)} -v_\a^{(k)} }, ~~~j=1,\dots,M_\a.
\label{bae1}
\end{eqnarray}
For a given Bethe root, the eigenstate is given by 
\begin{eqnarray}
\left|\{v_\a^{(j)} \}\right> =\prod_{k=1}^{M_\a}\left(\frac{S_1^+}{v_\a^{(k)} +\epsilon}+\frac{S_2^+}{v_\a^{(k)} -\epsilon}\right)
\left|\phi(p_\a,q_\a)\right> .
\label{state1}
\end{eqnarray} 
Using this explicit form for the eigenstates we then deduce that 
\begin{eqnarray*}
C_1\left|\{v_\a^{(j)} \}\right> &=& \frac{(p_\a^2-1)}{4}\left|\{v_\a^{(j)} \}\right> \\
C_2\left|\{v_\a^{(j)} \}\right> &=& \frac{(q_\a^2-1)}{4}\left|\{v_\a^{(j)} \}\right> \\
(S^z_1+S^z_2)\left|\{v_\a^{(j)} \}\right> &=& \left(M_\a+1+\frac{p_\a+q_\a}{2}\right)\left|\{v_\a^{(j)} \}\right> \\
K_1K_2\left|\{v_\a^{(j)} \}\right> &=& \frac{p_\a q_\a}{4}\left|\{v_\a^{(j)} \}\right> .
\end{eqnarray*}
Hence the energy of the Hamiltonian (\ref{ham1}) in terms of the Bethe roots $\{v_\a^{(j)} \}$ is 
\begin{eqnarray}
E=g(p_\a+M_\a)(q_\a+M_\a)+(p_\a-q_\a)\epsilon  -2\sum_{j=1}^{M_\a} v_\a^{(j)}  .
\label{nrg1}
\end{eqnarray}

For the representation (\ref{ham2}) the energies 
of the Hamiltonian $H_{\b}$ are as follows, 
\begin{eqnarray}
E_{\b}=(p_\b-q_\b)\epsilon  -2\sum_{j=1}^{M_\b} v_\b^{(j)}  
\end{eqnarray}
where the $\{v_\b^{(j)}  \}$ are solutions of the BAE
\begin{eqnarray}
\frac{2}{g}+\frac{q_\b}{v_\b^{(j)}  +\epsilon}+\frac{p_\b}{v_\b^{(j)}  -\epsilon}=\sum_{k\neq j}^{M_\b}\frac{2}{v_\b^{(j)}  -v_\b^{(k)}  }, 
~~~j=1,\dots,M_\b.
\label{bae2}
\end{eqnarray}
For a given solution, the eigenstate is given by 
\begin{eqnarray}
\left|\{v_\b^{(j)}  \}\right> =\prod_{k=1}^{M_\b}\left(\frac{S_3^+}{v_\b^{(k)}  +\epsilon}+\frac{S_4^+}{v_\b^{(k)}  -\epsilon}\right)
\left|\phi(p_\b,q_\b)\right> .
\label{state2}
\end{eqnarray} 
Note that if 
\begin{eqnarray*}
M_\b> \min(p_\b,q_\b),
\label{constraint}
\end{eqnarray*}
then (\ref{state2}) vanishes. This is in contrast to (\ref{state1}), for which there is no analogous constraint on $M_\a$. 
Taking the above form for the eigenstates, we find that 
\begin{eqnarray*}
C_3\left|\{v_\b^{(j)}  \}\right> &=& \frac{q_\b(q_\b+2)}{4}\left|\{v_\b^{(j)}  \}\right> \\
C_4\left|\{v_\b^{(j)}  \}\right> &=& \frac{p_\b(p_\b+2)}{4}\left|\{v_\b^{(j)}  \}\right> \\
(S^z_3+S^z_4)\left|\{v_\b^{(j)}  \}\right> &=& \left(M_\b-\frac{p_\b+q_\b}{2}\right)\left|\{v_\b^{(j)}  \}\right> \\
K_3K_4\left|\{v_\b^{(j)}  \}\right> &=& \frac{p_\b q_\b}{4}\left|\{v_\b^{(j)}  \}\right> .
\end{eqnarray*}
So that the corresponding energy for the Hamiltonian (\ref{ham2})  in terms of the Bethe roots $\{ v_\b^{(j)}   \}$ is  
\begin{eqnarray}
E=g(M_\b-p_\b)(M_\b-q_\b)-gM_\b+(p_\b-q_\b)\epsilon  -2\sum_{j=1}^{M_\b} v_\b^{(j)}   .
\label{nrg2}
\end{eqnarray}

In order to compare the two Bethe ansatz solutions we need to determine the relationship between the parameters
$\{p_\a,\,q_\a,\,M_\a\}$ and $\{p_\b,\,q_\b,\,M_\b\}$. {}From the form of the eigenstates (\ref{state1},\ref{state2}) it is deduced that
\begin{eqnarray*}
n_1+n_4 &=& q_\a+M_\a  =  q_\b, \\
n_2+n_3 &=& p_\a+M_\a  =  p_\b, \\
n_1-n_2 &=& -p_\a      =  M_\b-p_\b 
\end{eqnarray*} 
giving the solution 
\begin{eqnarray*}
q_\b&=& q_\a+M_\a,   \\ 
p_\b&=& p_\a+M_\a,   \\ 
M_\b&=& M_\a.    
\end{eqnarray*} 
Hence for every solution of (\ref{bae1}) with $\{p_\a,\,q_\a,\,M_\a\}$ giving energy $E$ via (\ref{nrg1}) there is a solution of 
(\ref{bae2}) with $\{p_\b=p_\a+M_\a,\,q_\b=q_\a+M_\a,\,M_\b=M_\a\}$ giving the same energy $E$ via (\ref{nrg2}). 
Note that in this correspondence the constraint (\ref{constraint}) is never violated. 
Our next goal is to use this result to establish a spectral equivalence in the QES sector of a Schr\"odinger equation.


\subsection{Equivalences of QES sectors} 


We start with the general form of Bethe ansatz equations 
\begin{eqnarray}
A+\frac{B}{v^{(j)} +\gamma/2}+\frac{C}{v^{(j)} -\gamma/2}=\sum_{k\neq j}^M \frac{2}{ v^{(j)} -v^{(k)} }
\label{bae3}
\end{eqnarray}
for $A,~B,~C\in \mathbb R $ and set 
\begin{eqnarray}
\psi(x)=(\cosh(x)-1)^{-(B/2+1/4)}(\cosh(x)+1)^{-(C/2+1/4)} \nn \\ 
\qquad\qquad\quad\times
\exp\left(\frac{A\gamma}{4}\cosh(x)\right)\prod_{j=1}^M\left(\frac{\gamma}{2}\cosh(x)+v^{(j)}
\right).
\label{psi2}
\end{eqnarray}
It can be shown \cite{uly} that $\psi(x)$ satisfies the Schr\"odinger equation 
\begin{eqnarray}
-\frac{d^2 \psi}{d x^2}+V(x)\psi={\mathcal{E}}\psi
\label{ode}
\end{eqnarray} 
where 
\begin{eqnarray*}
&&V(x;A,B,C,\gamma)\\
&&~~~~~~=M\left(M-B-C+\frac{A\gamma}{2}\cosh(x)-1\right) +\frac{1}{4}(B+C+1)^2 \\
&&~~~~~~~~~~~+\frac{A^2\gamma^2}{16}\sinh^2(x)+\frac{A\gamma(C-B)}{4}-\frac{A\gamma(B+C)}{4} \cosh(x) \\
&&~~~~~~~~~~~+\frac{(2B+1)(2B+3)}{8(\cosh(x)-1)}
-\frac{(2C+1)(2C+3)}{8(\cosh(x)+1)} \\
&{\mathcal{E}}=& -A\sum_{j=1}^M v^{(j)} . 
\end{eqnarray*} 
We note that this potential has the symmetry
\begin{eqnarray}
V(x;A,B,C,\gamma) = V(x+i\pi;A,C,B,-\gamma) .
\label{herm-pt}
\end{eqnarray}
Assuming $A\gamma$ is negative, we see that there is a spectral equivalence between the Hermitian problem with potential 
$V(x;A,B,C,\gamma)$ and the $PT$-symmetric problem $V(x;A,C,B,-\gamma)$ defined on the contour ${\Im}(x)=\pi$. 
Specifically, there is no $PT$-symmetry breaking in the latter.  
 
Returning to the Hamiltonian (\ref{ham3}), we fix $g=1$. Because the Bethe ansatz equations 
(\ref{bae1},\ref{bae2}) are of the same form as (\ref{bae3}), we can map the spectrum of (\ref{ham3}) 
to that of the QES sector of (\ref{ode})  by adding the appropriate terms to the potential. 
Since there are two Bethe ansatz solutions for (\ref{ham3}) we obtain the potentials
\begin{eqnarray*}
V_{\a}(x;p_\a,q_\a)&=& V(x;2,-(p_\a+1),-(q_\a+1),2\epsilon)+(p_\a+M_\a)(q_\a+M_\a)\\
&&~~~~~~~+(p_\a-q_\a)\epsilon,  \\
V_{\b}(x;p_\b,q_\b)&=& V(x;2,q_\b,p_\b,2\epsilon)+(M_\b-p_\b)(M_\b-q_\b)-M_\b\\
&&~~~~~~~+(p_\b-q_\b)\epsilon ,   
\end{eqnarray*} 
where for $V_{\a}$, $\displaystyle{\mathcal{E}}= -2\sum_{j=1}^{M_\a} v_\a^{(j)} $ while for $V_{\b}$, 
$\displaystyle{\mathcal{E}}= -2\sum_{j=1}^{M_\b} v_\b^{(j)} $ . 
We have already seen that the Bethe ansatz solutions for (\ref{ham3})
are equivalent when $M_\a = M_\b, ~p_\b=p_\a+M_\a,\,q_\b=q_\a+M_\a$. 
It follows that the potentials $V_{\a}(x;p_\a,q_\a)$ and
$V_{\b}(x;(p_\a+M_\a),(q_\a+M_\a))$ have the same QES spectrum.

However, the QES
wavefunctions of the Schr\"odinger equation (\ref{ode}) are not always
normalisable on the full real line.  The potential $V(x;A,B,C,\gamma)$
has a singularity at the origin whenever $(2B+1) (2B+3) \ne 0$.  When $B=-3/2$
or $-1/2$,  the potential is nonsingular and the QES wavefunctions
 (\ref{psi2}) (assuming $A\gamma$ is negative) can be extended to
 normalisable odd/even  wavefunctions 
 respectively   on the full real line~\cite{GLKO}.   

  It is
 interesting to consider the two nonsingular cases.  The above result 
 with 
 $(p_\a,q_\a)=(-1/2 , -3/2-M)$ establishes a QES spectral equivalence 
between a potential with  $B=-3/2$ and a potential with $B=-1/2$.  
 By adding  constant shifts to
  $V_{\a}(x;p_\a,q_\a)$ and
$V_{\b}(x;(p_\a+M_\a),(q_\a+M_\a))$, we can {\it  prove}  that 
 the  spectral equivalence extends to the full spectrum, except
 for the presence of  
a single $E=0$ energy level in the former.   In fact, the potentials are 
  supersymmetric partners~\cite{Wit,Shif}.  Set 
\[
 {\cal
  Q}_{\pm} (x)= \pm  \frac{d}{dx}+\frac{(M+1) 
\sinh x }{ 2( \cosh x+1)} +  \gamma \sinh (x) ,
\]
then 
\[
\fl  {\cal Q}_{+}{\cal Q}_{-} \psi(x) = \lf [ -\frac{d^2}{dx^2} +
V(x;2,-3/2,M-1/2,\gamma) -  \gamma (M+1) \ri ] \psi
(x)   =E \psi(x), 
\]
with corresponding $E=0$ 
eigenfunction 
\[
\psi(x)= (\cosh x+1)^{(M+1)/2} \times \exp \lf ( {\frac{ 
    \gamma}{2} \cosh x } \ri). 
\]
We immediately deduce that the supersymmetric partner 
\[
{\cal Q}_{-}{\cal Q}_{+} \Psi(x) =  E \Psi(x) 
\] 
has  potential
 $V(x;2,-1/2,M+1/2,\gamma)+M-\gamma (M+1)$. We have therefore  
 established complete  
 isospectrality between these Schr\"odinger problems, up to the $E=0$
 energy level.   Moreover, this spectral equivalence holds for all real 
values of $M$.

Finally, we remark that the unitary transformation (\ref{ut}) combined with the change of variable (\ref{cov1}) 
leaves the Hamiltonian invariant. Observe that (\ref{ut}) applied to (\ref{vac2}) has the effect that 
\begin{eqnarray}
p \longleftrightarrow q. 
\label{cov2}
\end{eqnarray}   
This is reflected in the symmetry of the Bethe ansatz equations (\ref{bae1},\ref{bae2}) which are invariant under
the combination of (\ref{cov1},\ref{cov2}). Thus the unitary transformation (\ref{ut}) effectively interchanges 
the Hermitian and 
$PT$-symmetric versions of the potentials $V_{\a}(x;p_\a,q_\a),\,V_{\b}(x;p_\b,q_\b)$, which explains the existence of the 
symmetry (\ref{herm-pt}). 

This scenario is somewhat different to the case of the QES sextic
potential discussed previously.  
There, the unitary transformation maps the QES spectrum into the
negative of the  
QES spectrum. The equivalence between the Hermitian and $PT$-symmetric
problems is  
due to the equivalence of the Bethe ansatz solutions. For the above
case, the unitary  
transformation maps between the Hermitian and $PT$-symmetric cases,
while the equivalence of the  
Bethe ansatz solutions gives a spectral equivalence between two
Hermitian QES potentials.

\section*{Acknowledgements}

TCD is partially supported by grant NAL/32601 from   the Nuffield Foundation.
KEH and JRL are funded by the
Australian Research Council through Discovery Project DP0557949.
KEH also acknowledges financial support from a University of Queensland 
Postdoctoral Research Fellowship for Women.

\end{document}